\documentclass[12pt,a4paper]{article}
\usepackage[DIV13]{typearea}
\usepackage{amsmath,amssymb,amsfonts}
\usepackage{color}
\usepackage[colorlinks]{hyperref}
\usepackage{graphicx} 

\newcommand{\eq}[1]{Eq.~\ref{#1}}
\newcommand{\Figref}[1]{Fig.~\ref{#1}}

\newcommand{\be}{\begin{equation}}
\newcommand{\ee}{\end{equation}}
\def\bea{\begin{eqnarray} }
\def\eea{ \end{eqnarray} } 
\newcommand{\tev}{\,\, \mathrm{TeV}}
\newcommand{\gev}{\,\, \mathrm{GeV}}

\begin{document}


\begin{center}
  {\Large\bf Stop-mass prediction in naturalness scenarios within MSSM-25\\ }
  \vspace{1cm}
  \renewcommand{\thefootnote}{\arabic{footnote}}
  \textbf{Shehu S. AbdusSalam\footnote[1]{Email:
      \texttt{Shehu.AbdusSalam@roma1.infn.it}}$^{(a),(b)}$ 
  }
  \\[5mm]
\end{center}
\textit{\small 
  $^{(a)}$INFN, University of Rome ``La Sapienza'', Piazzale A. Moro 2, I-00185 Roma, Italy\\[2mm]
  $^{(b)}$The Abdus Salam International Centre for Theoretical Physics, Strada Costiera 11, I-34151 Trieste, Italy\\[2mm]
}

\vspace{1cm}

\begin{abstract}
The ``top-down'' approach to minimal supersymmetric standard model
(MSSM) phenomenology provides model-dependent indications for a
natural stop mass. The approach is based on specific assumptions
about the supersymmetry-breaking energy scale and parameters
degeneracies. In order to
determine robust predictions we update the stop-mass 
prediction within the MSSM with 25 parameters (MSSM-25) by including
electroweak fine-tuning as ``naturalness data'' during the Bayesian
fits of the parameters to experimental data. The approximately
prior-independent results show that imposing naturalness, taken here
to mean a 25\% to 100\% fine-tuning, predicts a $1-2 \tev$ stop
mass. The posterior distributions for the neutralino-proton cross
sections indicate better prospects for probing the associated
neutralino cold dark matter (CDM) with future upgrades of the
detection facilities.  
\end{abstract}


\paragraph{Introduction:} 
Fitting the R-parity conserving MSSM-25 to pre-LHC experimental data
predicts a now discovered~\cite{Aad:2012tfa, Chatrchyan:2012ufa},
Higgs boson mass and the, so far undiscovered, stop mass to be
respectively $117 - 129 \gev$ and $2 - 3 \tev$ at 95\% Bayesian
credibility region~\cite{AbdusSalam:2008uv, AbdusSalam:2009qd}. The
SUSY-breaking parameters were simultaneously varied at the TeV-scale,
which the LHC 
is meant to probe, independent of hidden-sector physics, mediation
mechanisms and with minimal renormalisation group running
restrictions. In this article, we update the analysis in
Refs.~\cite{AbdusSalam:2008uv, AbdusSalam:2009qd}, within our MSSM-25
programme~\cite{AbdusSalam:2008uv, AbdusSalam:2009qd,
  AbdusSalam:2010qp, AbdusSalam:2011hd, AbdusSalam:2011fc,
  AbdusSalam:2012sy, AbdusSalam:2012ir}, for addressing the
implications of naturalness requirement on the stop mass
prediction. 

The studies on MSSM naturalness and its implication
on the stop mass can be classified into two main groups. On the one
hand, is the ``top-down'' approach which depends on various
well-motivated but ad-hoc simplifications of the MSSM parameters (such
as CMSSM, the constrained MSSM) or ad-hoc simplifications 
of the MSSM sparticle spectrum (the so-called simplified models)
as, for instance, in Refs.~\cite{deCarlos:1993yy, Chan:1997bi,
  Dimopoulos:1995mi, Berezinsky:1995cj, Cassel:2011tg,
  Ghilencea:2012qk, Kobayashi:2006fh, Arvanitaki:2012ps, ArkaniHamed:2012gw, Feng:2013pwa}. 
For this group, naturalness predicts that the lighter stop mass
$m_{\tilde t_1}$ can be $m_{\tilde t_1} \lesssim 173 
\gev$ or $700 \gev \lesssim m_{\tilde t_1}$ depending on whether the
unification scale, $\Lambda_{GUT}$, is $\Lambda_{GUT} \sim 10^{16}
\gev$ or $\Lambda_{GUT} \sim 10 \tev$ respectively. 
The ``bottom-up'' approach, on the other hand, depends solely on the
full MSSM sparticle spectrum and couplings specified at the
electroweak scale~\cite{Baer:2012up, Baer:2012cf, Baer:2012mv}. Both 
approaches involve either randomly generated ``predictions'' or ad-hoc 
parameters/spectrum simplifications for the effect of naturalness
requirement on the stop mass. 

In this article, we require that naturalness prediction for the stop
mass should be
assessed via unambiguous phenomenological frame and assumptions. For
this purpose, so far, we find the Bayesian approach such as done in 
Refs.~\cite{AbdusSalam:2008uv, AbdusSalam:2009qd} to be the most
appropriate tool since it has a systematic procedure for checking the
stability of conclusions with respect to the strength of data and
assumptions. The electroweak fine-tuning (EWFT) measure, 
$\Delta_{EW}$, defined in Ref.~\cite{Baer:2012up}, which provides a
measure of fine-tuning given a full MSSM spectrum, and couplings at the
electroweak scale and Bayesian techniques will be used
for finding robust indication for natural $m_{\tilde t_1}$ within
MSSM-25. We consider $\Delta_{EW}$ as the most
appropriate for our analysis given the match between its construction
and the MSSM-25 parameterisation procedures. Both were developed as
purely electroweak scale-based phenomena independent of SUSY-breaking
physics and parameters renormalisation group running. The measures
defined in Refs.~\cite{Nucl.Phys.B306.63, ellis} depend on the
parameters and the restrictions with which they usually come. The
$\Delta_{EW}$ naturalness requirement cuts will be imposed as a
constraint while exploring the MSSM-25 parameter space unlike via the
marginalisation procedure as in Refs.~\cite{Nucl.Phys.B306.63, ellis,
  Cabrera:2008tj, Ghilencea:2013nxa}.

\paragraph{The electroweak fine-tuning measure:}
The fine-tuning measure is based on the minimisation of the 1-loop 
corrected potential energy, $V + \Delta V$, of the Higgs boson fields
which leads to  
\be \label{Hmin1} \frac{m_Z^2}{2} = \frac{m_{H_d}^2 + \Sigma_d^d - 
    (m_{H_u}^2+\Sigma_u^u)\tan^2\beta}{\tan^2\beta -1} -\mu^2\ee
where $\Sigma_u^u$ and $\Sigma_d^d$ are radiative corrections that
arise from the derivatives of $\Delta V$ evaluated at the
minimum~\cite{Baer:2012up, Baer:2012cf}. $m_{H_u}$ and $m_{H_d}$ are
respectively the up-type and down-type Higgs doublet mass parameters,
$\tan \beta=\left<H_d\right>/\left<H_u\right>$ is the ratio of their
vacuum expectation values. $\mu$ represents the Higgs doublets mixing 
parameter. Naturalness requires each term in the right hand side of
\eq{Hmin1} to be comparable to $m^2_Z/2.$ 
The definition 
\be \label{DEW} \Delta_{EW} \equiv max_i \left(C_i\right) /
(m_Z^2/2) \ee
accommodates the fact that for obtaining a natural value of
$m_Z$ then the terms $C_i$, with $i=H_d,\ H_u$, $\mu$,
$\Sigma_u^u(k)$, $\Sigma_d^d(k)$, where $k$ denotes the various
particles and sparticles contributions, must have an order $m_Z^2/2$
absolute values. We use only the contributions (the case $i =
\tilde{t}_{1,2}, \tilde{b}_{1,2}$) from terms that
couple the most to the Higgs sector:
\be \begin{split} 
  C_\mu &= |-\mu^2|, \\
  C_{H_u} &= |-m_{H_u}^2\tan^2\beta /(\tan^2\beta -1)|, \,\\
  C_{H_d} &= |m_{H_d}^2/(\tan^2\beta -1)|,\\
  C_{\Sigma^d_d} &= | \Sigma^d_d/(\tan^2\beta -1)|, \\
  C_{\Sigma^u_u} &= | -\Sigma^u_u \, \tan^2 \beta/(\tan^2\beta -1)|, \\
  \Sigma^{d,u}_{d,u} &= \Sigma_{i} \, | \Sigma^{d,u}_{d,u} (i)|.
\end{split} 
\ee 
The expressions for $\Sigma^{d,u}_{d,u} (i)$ are given in the Appendix.

\paragraph{Bayesian global fit method and naturalness:}  \label{bayesm}
The cut applied on the EWFT measure \eq{DEW} to restrict its values to
acceptable level were considered as ``naturalness data'' for the
Bayesian fits (with linear and logarithmic priors on the parameters)
of the MSSM-25 to data (listed in \eq{alldata}).  
Any prior-independent result~\footnote{If the posterior distribution
  of a quantity (for example the stop 
mass) coming from a fit with a flat prior on the parameters is widely
different from that coming from a fit with a logarithmic prior then
the result (the stop mass distribution) is said to be
prior-dependent. Otherwise the result may be considered to be approximately
or fully prior-independent.} or posterior distribution
concerning the stop mass obtained from the Bayesian fits will
constitute a 
robust ``naturalness prediction'' within the MSSM-25. For
completeness, we 
shall introduce Bayes' theorem first before describing the procedure
for the fits. Bayes' theorem states that
given a model hypothesis, $\cal{H}$, with parameters, $\underline
\theta$, and a set of data, $\underline d$, for constraining the model
then  
\be \label{bayes} 
p(\underline \theta |\underline d, {\cal{H}}) = \frac{p(\underline
  d|\underline \theta,{\cal{H}}) p(\underline
  \theta|{\cal{H}})}{p(\underline d|{\cal{H}})}  
\ee where $p(\underline \theta | {\cal{H}})$ is the prior probability
distribution which provides information about the model parameters
before the data $\underline d$ is taken into consideration.  
The information about the model in light of the data is 
represented by the probability distribution $p(\underline
d|\underline \theta, {\cal{H}})$ or the likelihood of the model given
the data. $p(\underline \theta| \underline d,{\cal{H}})$ is the 
probability distribution of the model parameters given the
data. 

{\it The Bayesian fit procedure:} \label{ftexplore}
The 
 fits were done within the context, ${\cal H}$, that
Nature is supersymmetric as captured by MSSM-25 and that the
neutralino lightest supersymmetric particle (LSP) explains at least
partially the observed cold dark matter (CDM) relic
density~\cite{AbdusSalam:2010qp}. The MSSM-25 parameters are
\be \label{20par}
\underline{\theta} = \{ M_{1,2,3};\;\; m^{3rd \, gen}_{\tilde{f}_{Q,U,D,L,E}},\;\; 
m^{1st/2nd \, gen}_{\tilde{f}_{Q,U,D,L,E}}; \;\;A_{t,b,\tau,\mu=e},
\;\;m^2_{H_{u,d}}, \;\;\tan \beta; \;\; m_Z, \;\;m_t, \;\;m_b,
\;\;\alpha_{em}^{-1}, \;\;\alpha_s \}
\ee
where $M_1$, $M_2$ and $M_3$ are the gaugino mass parameters (allowed
in the range -4 to 4 TeV) and 
$m_{\tilde f}$ are the sfermion mass parameters (allowed in the range
100 GeV to 4 TeV). $A_{t,b,\tau,\mu=e} \in [-8, 8]$ TeV
represents the trilinear scalar couplings, while the Higgs-sector
parameters are specified by the two Higgs doublet masses $m^2_{H_1}$,
$m^2_{H_2}$ (with $m^2 \in sign(m)\,[-4, 4]^2 \textrm{ TeV}^2$ ), the
ratio of the vacuum expectation values $\tan 
\beta=\left<H_2\right>/\left<H_1\right>$ (allowed between 2 and 60) and
$sign(\mu)$ is the sign of the Higgs doublets mixing parameter
(allowed to be randomly 
$\pm 1.$)  The mass of the Z-boson, $m_Z = 91.1876 \pm 0.0021$; the
top quark mass, $m_t = 172.6 \pm 1.4 \gev$, the bottom quark mass, $m_b = 4.2
\pm 0.07 \gev$, the electromagnetic coupling, $\alpha_{em}^{-1} =
127.918 \pm 0.018$, and the strong interaction coupling, $\alpha_s = 
0.1172 \pm 0.002$, were all set to vary in a
Gaussian manner with central values and deviations according to the 
experimental results. The two Bayesian global fits were done with
linear and logarithmic prior assumptions for the SUSY-breaking 
parameters. 

Adding the EWFT measure $\Delta_{EW}$ to the list of data makes
\be \underline{d} = d_{pre-LHC} \, \oplus \,
d_{\Delta_{EW}} = \{ \mu_i, \sigma_i \} \, \oplus \{\Delta_{EW}^{-1} \geq
5\%\}. \ee
Here $d_{pre-LHC}$~\footnote{We are going to maintain this
  notation despite the fact that the Higgs boson mass is no longer
  a ``pre-LHC'' observable.} represents the Higgs boson mass, the
electroweak physics, 
B-physics and the cold dark matter relic density observables,
represented by their corresponding central values ($\mu_i$) and
errors ($\sigma_i$) where $i = 1,2,\, \ldots,$ enumerates the
observables as summarised in Table~\ref{tab:obs} and the list: 
\be
\begin{split} \label{alldata}
\underline O = &\{m_W,\; \sin^2\, \theta^{lep}_{eff},\; \Gamma_Z,\;
\delta 
a_{\mu},\; R_l^0,\; A_{fb}^{0,l},\; A^l = A^e,\; R_{b,c}^0,\;
A_{fb}^{b,c},\; A^{b,c},\;  BR(B \rightarrow X_s \, \gamma),\; \\
& BR(B_s\rightarrow \mu^+ \, \mu^-), \; \Delta_{0-},\; R_{BR(B_u
  \rightarrow \tau \nu)},\; R_{\Delta M_{B_s}}, \Omega_{CDM}h^2,\;
m_h, \, \Delta_{EW}^{-1} \geq 5\% \, \}. 
\end{split}
\ee 

The SUSY spectrum calculator {\sc SOFTSUSY}~\cite{Allanach:2001kg}
was used for computing the spectra, mixing angles and couplings from
the input soft 
SUSY-breaking parameters, {\sc micrOMEGAs}~\cite{Belanger:2008sj} for
computing the neutralino CDM relic density and the anomalous magnetic
moment of the muon $\delta a_\mu$, {\sc
  SuperIso}~\cite{Mahmoudi:2007vz} for predicting the branching ratios
$BR(B_s \rightarrow \mu^+ \mu^-)$, $BR(B \rightarrow s \gamma)$ and
the isospin asymmetry, $\Delta_{0-}$, in the decays $B \rightarrow K^*
\gamma$, and {\sc susyPOPE}~\cite{Heinemeyer:2006px,Heinemeyer:2007bw}
for computing precision observables that include the $W$-boson mass
$m_W$, the effective leptonic mixing angle variable $\sin^2 
\theta^{lep}_{eff}$, the total $Z$-boson decay width, $\Gamma_Z$, and
the other electroweak observables whose experimentally
determined central values and associated errors are summarised in
Table~\ref{tab:obs}. 
\begin{table} 
\begin{center}{\begin{tabular}{|cl||cl|}
\hline
Observable & Constraint & Observable & Constraint  \\ 
\hline
$m_W$ [GeV]& $80.399 \pm  0.027$ \cite{verzo}&$A^l = A^e$& $0.1513 \pm
0.0021$ \cite{:2005ema} \\
$\Gamma_Z$ [GeV]& $2.4952 \pm 0.0025$ \cite{:2005ema}&$A^b$ & $0.923
\pm 0.020$ \cite{:2005ema}\\ 
$\sin^2\, \theta_{eff}^{lep}$  & $0.2324 \pm 0.0012$ \cite{:2005ema}&$A^c$ & $0.670 \pm 0.027$ \cite{:2005ema}\\  
$\delta a_\mu $ & $(30.2 \pm 9.0) \times 10^{10}$
\cite{Bennett:2006fi,Davier:2007ua} &$Br(B\rightarrow
X_s \gamma)$ & $(3.55 \pm 0.42) \times 10^{4}$ \cite{Barberio:2007cr}\\  
$R_l^0$ & $20.767 \pm 0.025$ \cite{:2005ema} &$Br(B_s \rightarrow \mu^+ \mu^-)$ & $
3.2^{+1.5}_{-1.2} \times 10^{-9}$ \cite{Aaij:2012nna}\\  
$R_b^0$ & $0.21629 \pm 0.00066$ \cite{:2005ema}&$R_{\Delta M_{B_s}}$ & $0.85 \pm 0.11$\cite{Abulencia:2006ze}\\ 
$R_c^0$ & $0.1721 \pm 0.0030$ \cite{:2005ema}&$R_{Br(B_u \rightarrow \tau \nu)}$&
$1.26 \pm 0.41$ \cite{Aubert:2004kz,paoti,hep-lat/0507015}\\ 
$A_{\textrm{FB}}^b$ & $0.0992 \pm 0.0016$ \cite{:2005ema}&$\Delta_{0-}$ & $0.0375 \pm
0.0289$\cite{J.Phys.G33.1}\\  
$A_{\textrm{FB}}^c$ & $0.0707 \pm 0.035$ \cite{:2005ema}&$\Omega_{CDM} h^2$ & $0.11
\pm 0.02 $ \cite{0803.0547}\\ 
 & & $m_h$ & $125.6 \pm 3.0$  [GeV]\cite{ATLAS:2013mma, CMS:yva}\\ 
\hline
\end{tabular}}\end{center}
\caption{Summary for the central values and errors for the Higgs boson
  mass, the electroweak physics observables, B-physics observables and
  cold dark matter relic density constraints.\label{tab:obs}}  
\end{table}
%
%
The posterior probability is thus given by Bayes' theorem,
\eq{bayes}, as  
\be \label{posterior}
 p(\underline \theta |\underline d, {\cal{H}}) = L_{\Delta_{EW}} \, L_{CDM}(x) \prod_i \, \frac{
   e^{\left[- (O_i - \mu_i)^2/2 \sigma_i^2\right]}}{\sqrt{2\pi
     \sigma_i^2}} \, \frac{p(\underline \theta | {\cal{H}})}
 {p(\underline d | {\cal{H}})}; 
\ee

\be
L_{\Delta_{EW}} = 
\begin{cases}
1, & \textrm{if $\Delta_{EW}^{-1} \geq 5\%$} \\  
0, & \textrm{if $\Delta_{EW}^{-1} < 5\%$} \\
\end{cases},
\quad
L_{CDM}(x) = 
\begin{cases}
1/(y + \sqrt{\pi s^2/2}), & \textrm{if $x < y$} \\  
e^{\left[-(x-y)^2/2s^2\right]}/(y + \sqrt{\pi s^2/2}),
& \textrm{if $x \geq y$} \\
\end{cases},
\ee
where the index $i$ run over the different experimental observables
(data) other than the CDM relic density, $x$ represents the predicted
value of the neutralino CDM relic density, $y = 0.11$ is the WMAP central
value quoted in Table~\ref{tab:obs} and $s=0.02$ the inflated
error. The likelihood contribution coming from the CDM relic density
is given by $L_{CDM}(x)$ which is purely Gaussian when the predicted relic
density $x$ is greater than the experimental central value $y =
0.11$ thus imposing penalisation for CDM over-production. No
penalisation is imposed when $x < y$. This way the 
fits allow for the possibility for multicomponent CDM such that the
non-neutralino LSP component(s) will account for the relic density
deficit. We used the {\sc MultiNest}~\cite{Feroz:2007kg, Feroz:2008xx} 
package that implements nested sampling algorithm~\cite{Skilling} for
exploring the MSSM-25 parameters by including the requirement of a
minimal fine-tuning, $\Delta_{EW} \leq 20$, in the likelihood
function. 

Another set of fits were done with $\Delta_{EW} \leq 4$, besides
the $\Delta_{EW} \leq 20$ one, as the minimally possible fine-tuning
in the sense of the upper bound of $\Delta_{EW} = 2 \pm 2$. This is
for the purpose of checking the stability of the stop mass posterior
distribution with respect to tighter $\Delta_{EW}$ (less fine-tuning)
cuts.  The reason for considering $\Delta_{EW} = 2 \pm 2$ is as
follows. In 
principle, the model point with no fine-tuning at all, according to the
chosen measure, will have $\Delta_{EW} \rightarrow 0$. But such a model
point with infinitesimally small fine-tuning is difficult to obtain in
practice. Therefore we suppose or define $\Delta_{EW} = 2$ which
corresponds to a fine-tuning not worse than $50\%$ to be a reasonable
minimum within a $100\%$ (of $\Delta_{EW} = 2$) theoretical error
allowance i.e. $\delta \Delta_{EW} = \pm 2$. With these assumptions,
then $\Delta_{EW} \leq 4$ is practically the most strong naturalness
cut we could impose for the fits.  
%

\begin{figure}[!t]
  \centering
  \includegraphics[width=8cm]{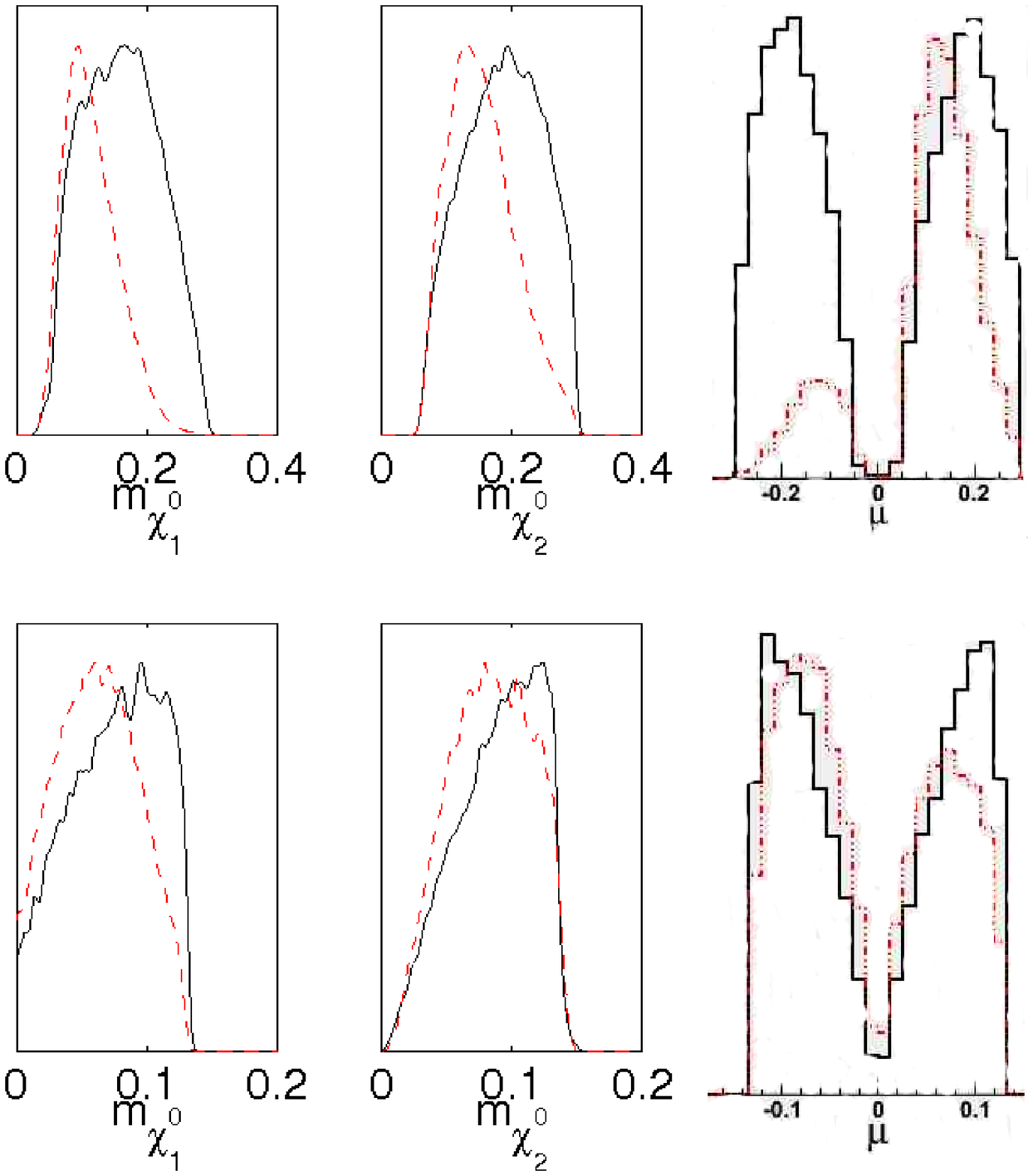}
  \includegraphics[width=8cm]{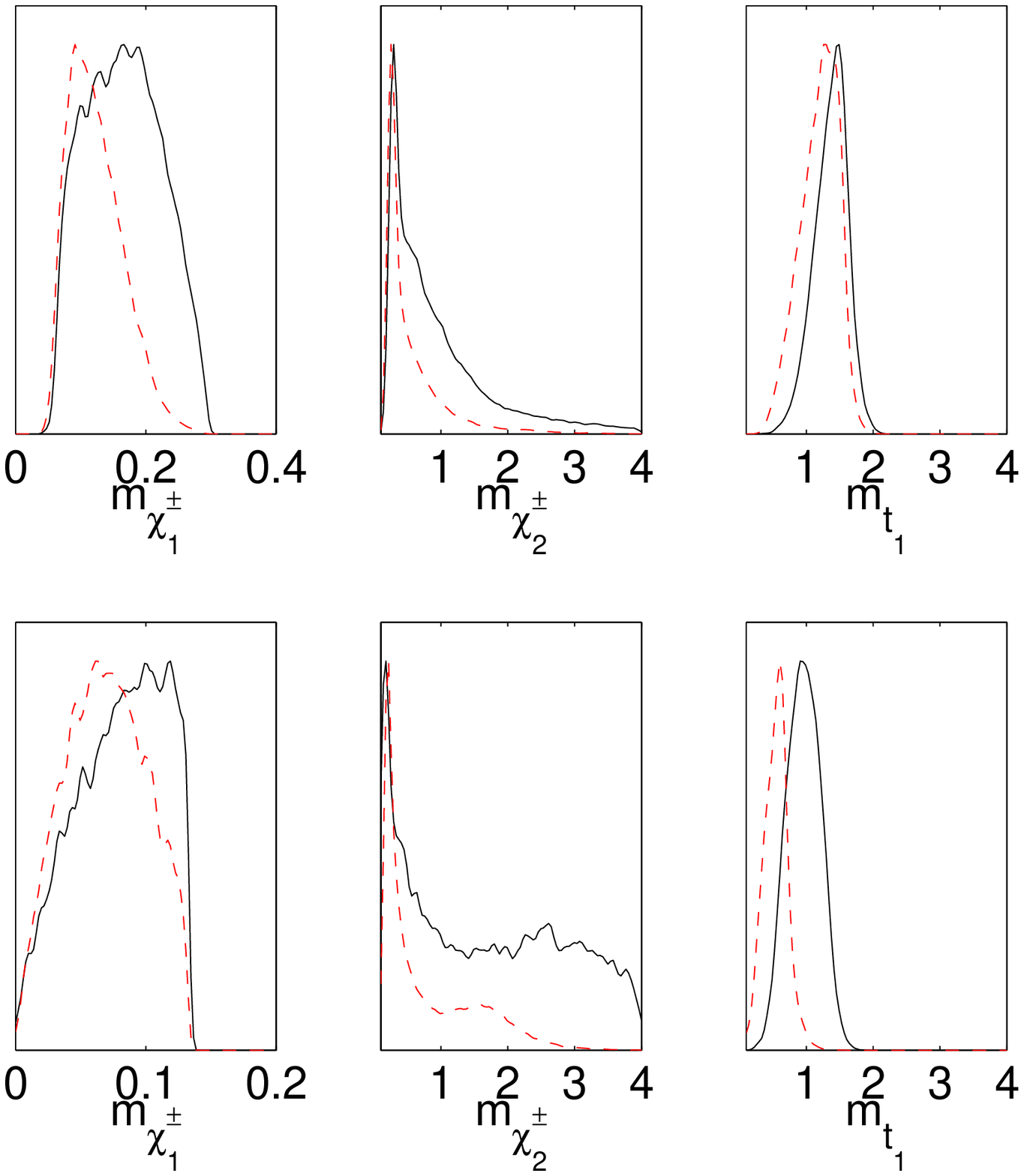}
  \caption{\small{{\bf (First row:)} The plots show the posterior
      distributions of sparticles with an approximately
      prior-independent feature over limited regions from the
      $\Delta_{EW} 
      \leq 20$ fit. Dotted(solid) curves are for the log(flat) prior
      fits of the MSSM-25. All the masses are in TeV units.
      {\bf (Second row:)} The plots are the same as in the first row
      but from the $\Delta_{EW} \leq 4$ fit. Note that the $m_{\tilde t_1}$
      distribution is now prior-dependent due to the relaxation of the
      Higgs-boson mass constraint. The $m_{\tilde \chi^0_{1,2}}$ and
      $m_{\tilde \chi^{\pm}_1}$ are now preferred to be lighter
      relative to the $\Delta_{EW} \leq 20$ results because of the
      tighter constraint on the Higgs bosons mixing parameter $\mu$ by
      imposing $\Delta_{EW} \leq 4$.}}  
  \label{masses}
\end{figure}

\begin{figure}[!t]
  \centering
  \includegraphics[width=8cm]{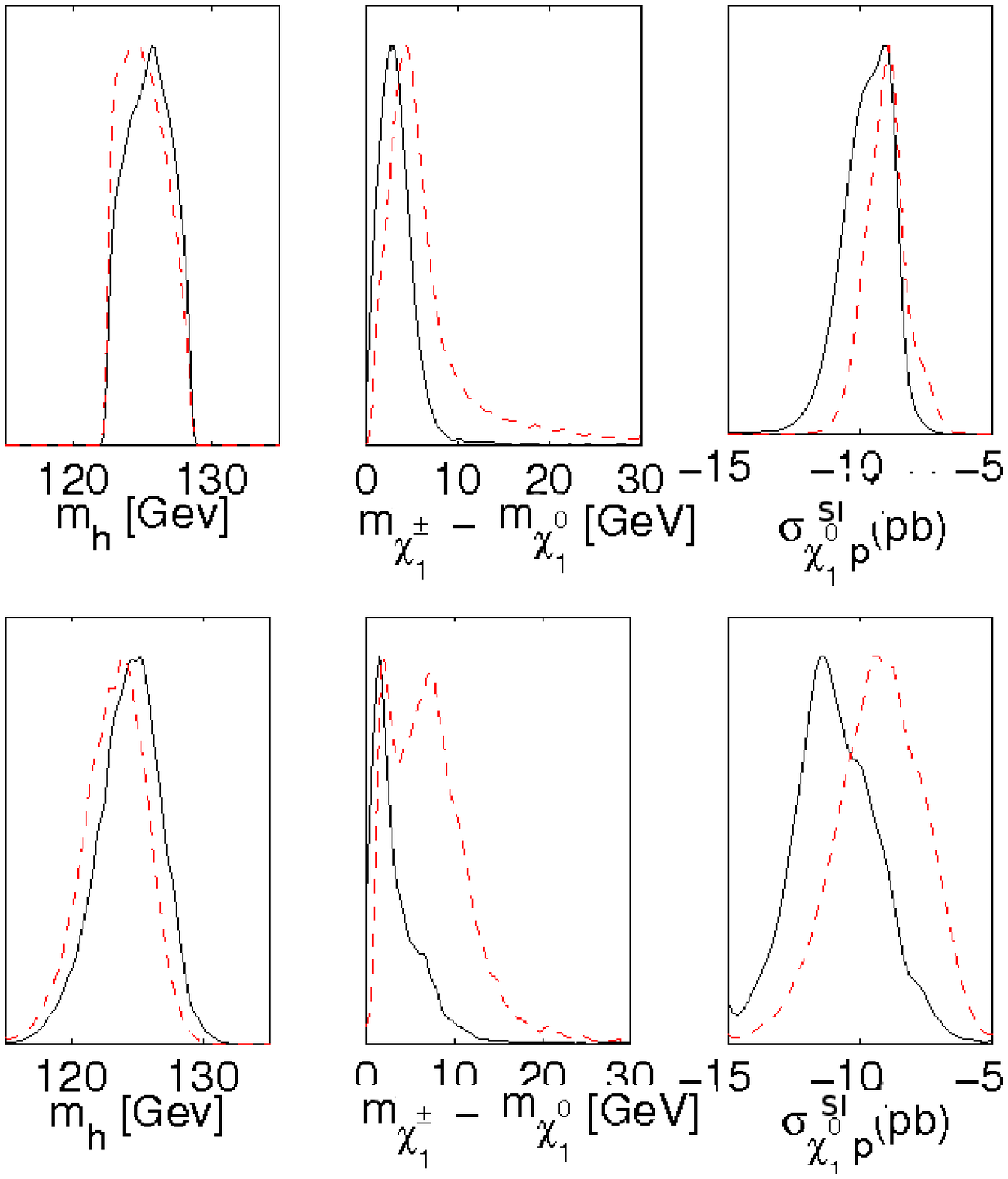}
  \includegraphics[width=8cm]{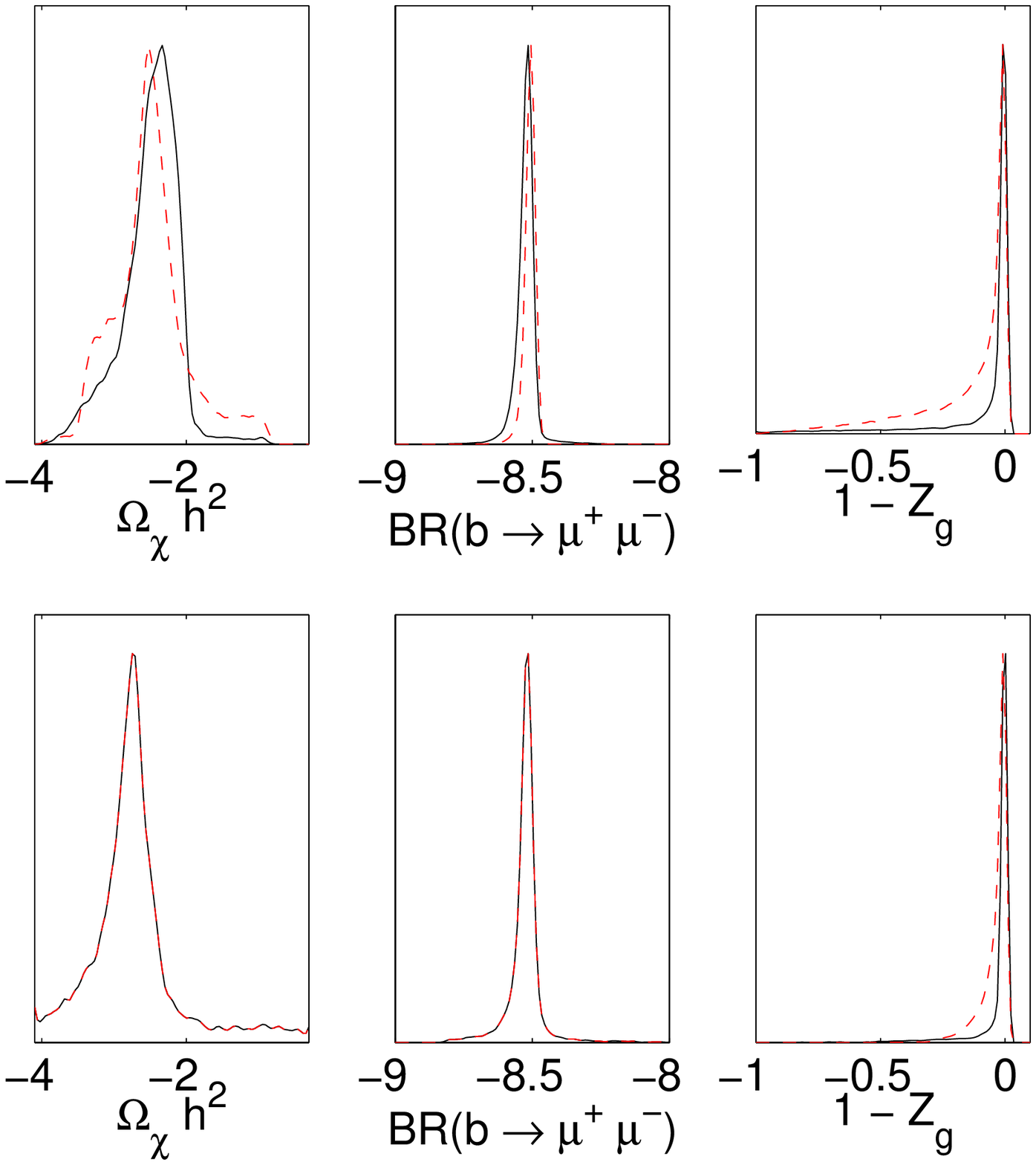}
  \caption{\small{{\bf (First row:)} The plots show the $\Delta_{EW}
      \leq 20$ posterior
      distributions for the Higgs boson mass $m_h$, the mass difference
      $\Delta m = m_{\chi_1^{\pm}} - m_{\chi_1^0}$, the spin-independent
      neutralino-proton cross section, $\sigma^{SI}_{\chi_1^0\,p}$, the
      neutralino relic density 
      $\Omega_{\chi^0_1}\,h^2$, the branching ratio of the decay $B_s
      \rightarrow \mu^+ \mu^-$, and the measure for quantifying the
      nature of the neutralino components $1-Z_g$ described in the
      text. All the x-axes other than the masses are in the log-10
      bases. {\bf (Second row:)} The plots are the same as in the
      first row but from the $\Delta_{EW} \leq 4$ fits.}}    
\label{observs}
\end{figure}

The $\Delta_{EW} \leq 4$ fits were done with a relaxed
Higgs boson mass constraints by allowing $m_h < 122 \gev$ but with  
Gaussian suppressed probabilities around $m_h = 125.6 \pm 3.0
\gev$. For 
the first set of fits with $\Delta_{EW} \leq 20$ all MSSM-25 points
with $m_h < 122 \gev$ were discarded. This little but significant
change on the Higgs boson mass constraint is used explicitly to point to
the effects of the fine-tuning versus $m_h$ cuts on the stop mass (see
the tails of the $m_h$ distributions in \Figref{observs} and the stop
mass distributions in \Figref{masses}.) 

\paragraph{Sparticle mass, $\tilde \chi^0_1$ relic density and $BR(B_s
  \rightarrow \mu^+ \mu^-)$ predictions:} \label{spredictions}
Most of the posterior distributions from both fits with
$\Delta_{EW} \leq 20$ and $\Delta_{EW} \leq 4$ have a
prior-dependent feature except, as is expected from \eq{Hmin1}, for
the neutralino, chargino and the lighter stops masses. The
approximately prior-independent tendency for $m_{\tilde t_1}$,
$m_{\tilde \chi^0_{1,2}}$ and $m_{\tilde \chi^{\pm}_1}$ distributions
can be seen in \Figref{masses}. It can be deduced from the results
that naturalness imposes upper bounds on $m_{\tilde \chi^0_{1,2}}$ and
$m_{\tilde \chi^{\pm}_1}$. This is because of the tree-level
fine-tuning restriction on the $\mu$ parameter to be near the Z-boson
mass, $m_Z$. The effect on $m_{\tilde t_1}$ and $m_{\tilde g}$ are at
the 1-loop and 2-loop levels respectively. It is the Higgs boson mass
constraint $m_h \sim 125 \gev$ that pulls the fits in fixing
$m_{\tilde t_1} \sim 1-2 \tev$ to be approximately prior independent
as shown on the first row of \Figref{masses}. Relaxing this constraint
has the effect of lowering the stop mass magnitudes as shown in the
second row of \Figref{masses}. Thus the $m_h \sim 125 \gev$ and the
EWFT constraint corner the stops to $m_{\tilde t_1} \sim 1-2
\tev$. This region is well 
above the ATLAS and CMS bounds in Refs.~\cite{ATLAS:2012rov, ATLAS:2013pla,
  Aad:2012ywa,Aad:2012xqa, Aad:2012tx, Aad:2012uu, Aad:2012yr,
  ATLAS:2013cma, TheATLAScollaboration:2013gha,
  TheATLAScollaboration:2013xha, TheATLAScollaboration:2013aia,
  Chatrchyan:2013lya, CMS:2013cfa, CMS:2013hda}. The bits of parameter
space with very light ($\Delta m \lesssim m_{\pi^{\pm}}$) to
intermediate ($m_{\pi^{\pm}} \lesssim \Delta m \lesssim 300$
MeV) chargino-neutralino mass difference $\Delta m =
m_{\chi_1^{\pm}} - m_{\chi_1^0}$ up to around $5 \gev$ (see
\Figref{observs} for the distributions of $\Delta m$) are constrained
by LEP and LHC results such as in
Refs.~\cite{Aad:2013yna,Heister:2002mn,Acciarri:2000wy,Abbiendi:2002vz}. 
However, the effect of the cuts on the posterior distribution is
insignificant.  

The neutralino mass-eigenstate is made of bino $\tilde{b}$, wino
$\tilde{w}^3$ and Higgsinos $\tilde{H}_{1,2}$ combination: 
\be \label{nis}
\tilde{\chi}_1^0 = N_{11}\tilde{b} + N_{12}\tilde{w}^3 +
   N_{13}\tilde{{{H}}_1^0} + N_{14}\tilde{H_2^0}, \quad \sum_{i=1,2,3,4}
   (N_{1i})^2 = 1
\ee
where $N_{1i}$ with $i=1,2,3,4$ are coefficient depending on soft-SUSY
breaking terms~\cite{ElKheishen:1992yv}. In \Figref{observs}, $(1 -
Z_g)$ where $Z_g = |N_{11}|^2 + |N_{22}|^2$ quantifies the nature of
the neutralino to be dominantly Higgsino- or gaugino-like for $Z_g$
approximately equal to unity or zero respectively. The
neutralino-chargino mass quasi-degeneracy (see the $\Delta m$ plots in
\Figref{observs}) together with the
Higgsino-nature of the neutralino enhances the primordial
co-annihilations of the neutralino CDM to the effect that the relic
density is much lower than the observed value around 0.1. The
remaining observed relic density has to be accounted for by a
non-neutralino CDM component(s). The posterior distribution for the
branching ratio of the decay $B_s \rightarrow \mu^+ \mu^-$ is also
approximately prior-independent near the standard model
value~\footnote{We shall analyse this decay in a separate report.}.

\begin{figure}[!t]
\includegraphics[width=8cm]{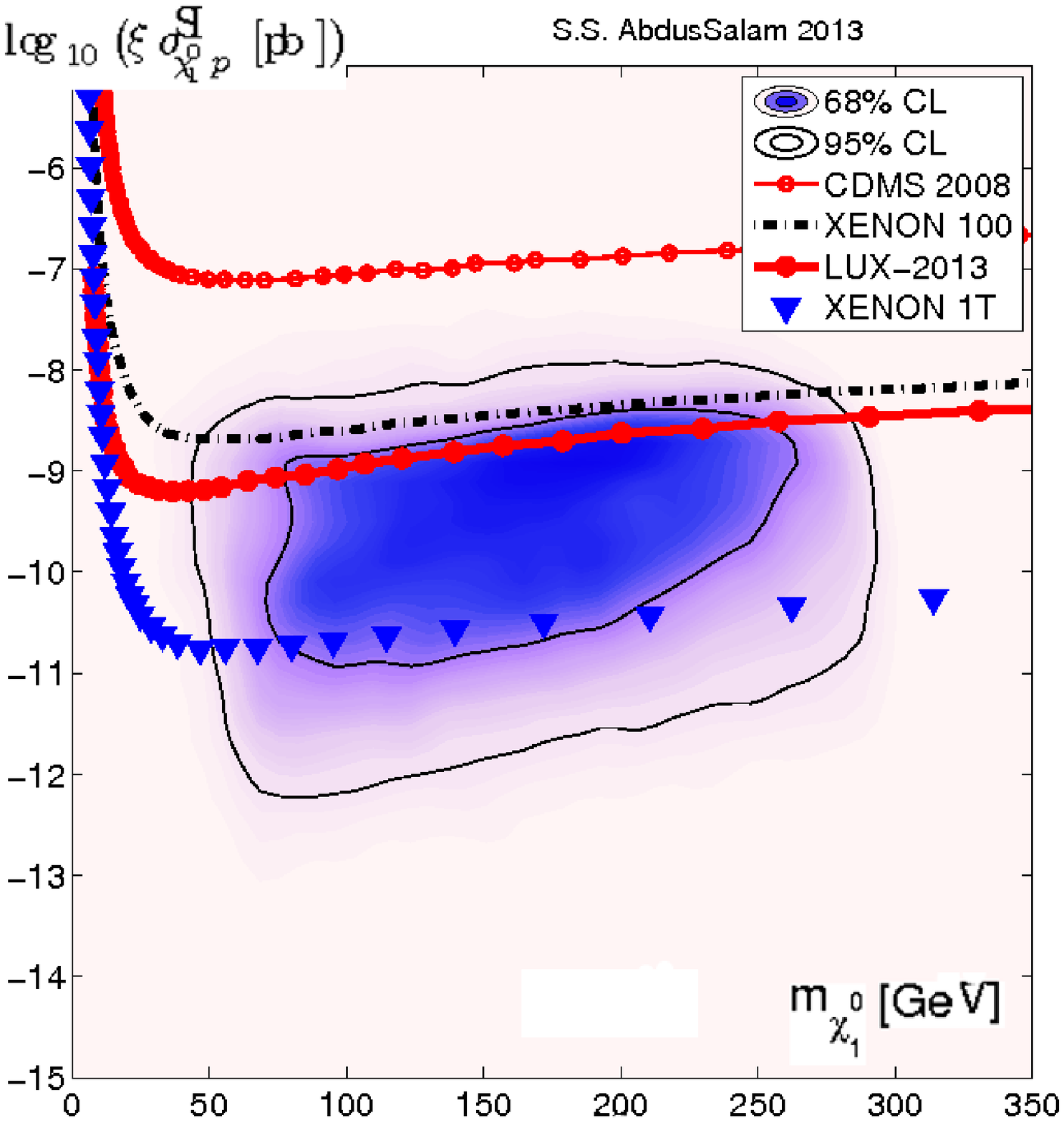}
\includegraphics[width=8cm]{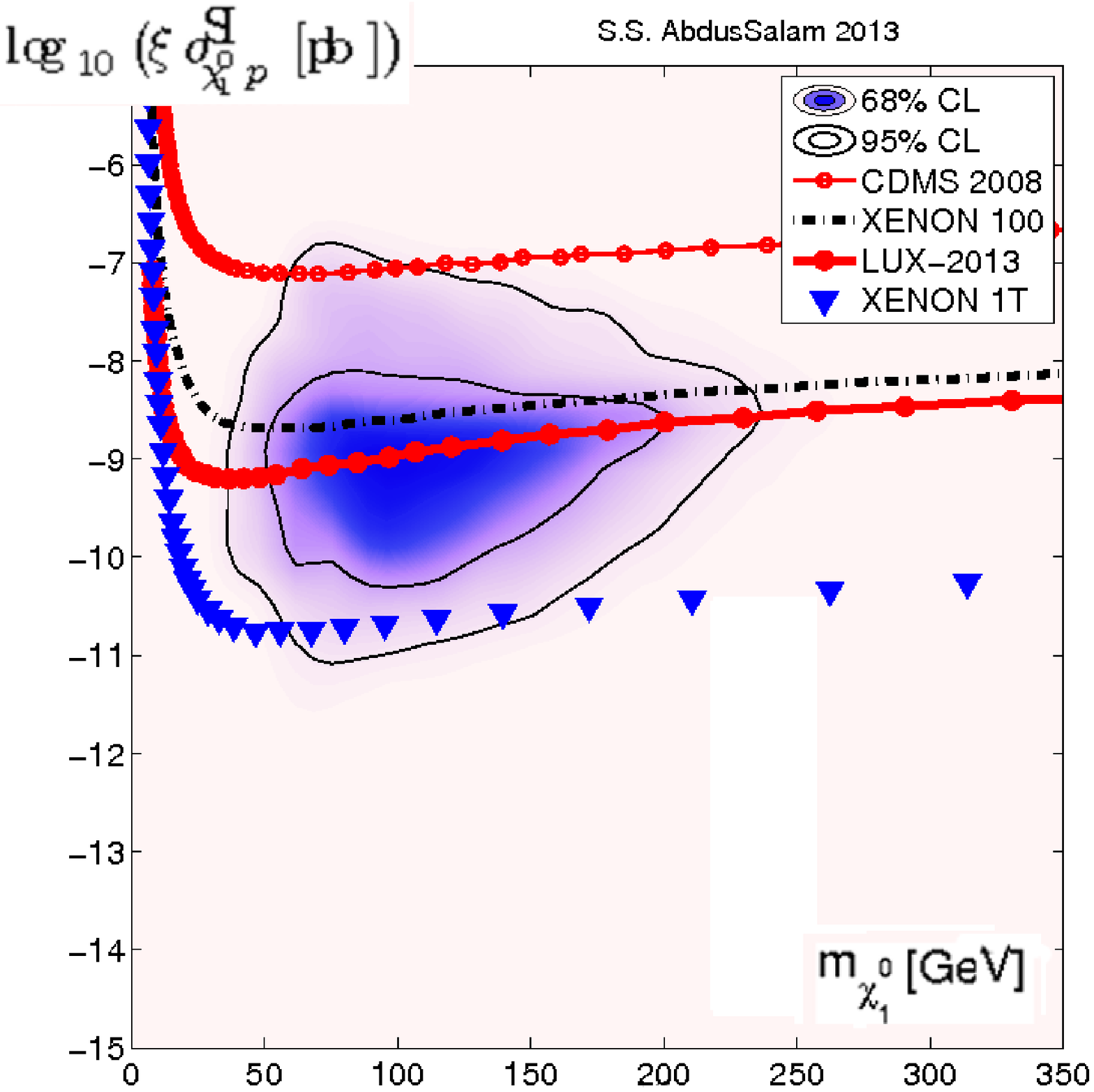}
  \caption{\small{The plots show the posterior distributions of the
      neutralino-proton spin-independent scattering cross section
      $\sigma^{SI}_{\chi_1^0\,p}$ (reduced by the factor $\xi =
      \Omega_{CDM} h^2/0.11$ described in the text) for the MSSM-25 with 
    linear (left) and log (right) prior measures. The CDMS-2008 $90\%$
    confidence level~\cite{Akerib:2005za},
    XENON-100~\cite{Aprile:2012nq}, LUX-2013~\cite{Akerib:2013tjd} and
    the future-projected XENON-1TN~\cite{Aprile:2012zx} upper bound constraints
    are shown as summarised by the legend of the plots. The solid contour
    lines show the $68\%$ and $95\%$ Bayesian credibility regions.}}
\label{sigmap}
\end{figure}

\paragraph{Prospects for direct detection of the $\tilde \chi^0_1$ CDM:}
Here we address the prospects of direct detection of the
neutralino part of CDM. The spin-independent neutralino-proton cross
section, $\sigma^{SI}_{\chi_1^0\,p}$, 
distributions against the neutralino mass are shown in
\Figref{sigmap}. In order to account for the fact that the
neutralino makes up only part of the dark matter relics, the cross
sections are rescaled by a factor $\xi = \Omega_{\chi_1^0} h^2/0.11$
as suggested in Ref.~\cite{Bottino:2000jx}. It can be seen from 
\Figref{sigmap} that the past/current dark matter direct detection
results such as from CDMS~\cite{Akerib:2005za} and
XENON~\cite{Aprile:2012nq} collaborations will hardly probe the
interesting regions within the MSSM-25. The result from the LUX
experimental collaboration~\cite{Akerib:2013tjd} is a bit different
given that the 90\% confidence limit on the cross section crosses over
a small but relatively more significant probability mass of the favoured
region compared to the earlier bounds. The prospect for detection is
much enhanced for future upgrades of the detectors such as the
ton-scale liquid Xenon detector~\cite{Aprile:2012zx}. 

\paragraph{Conclusions:}  \label{roundup}
We have addressed the question of SUSY naturalness within R-parity
conserving MSSM-25. We use naturalness requirements as a ``fine-tuning
data'' on the same footing as, for example, the CDM relic density data
in a Bayesian statistical method for fitting the MSSM-25
parameters. Doing this allows for a robust assessment of naturalness  
criterion strength in constraining the SUSY parameters. Two separate
fittings were performed with naturalness data $\Delta_{EW} \leq 20$
and $\Delta_{EW} \leq 4$. The results show that the combination of
Higgs-boson mass constraint and naturalness requirements corners the
stop mass to $1-2 \tev$. This reduction relative to the pre-LHC
prediction of $2-3 \tev$~\cite{AbdusSalam:2009qd} is due to the
naturalness pull. The stop prediction is associated with an
upper bounds on $m_{\tilde \chi^0_{1,2}}$ and $m_{\tilde
  \chi^{\pm}_1}$ to be sub-half TeV. The LUX-2013 experiment is
beginning to probe slightly the favoured region of parameter
space. The prospects for $\tilde \chi^0_1$ CDM discovery look better
with future upgrades of the direct detection facilities.

\paragraph{Acknowledgement:}
The author acknowledges useful discussions with H.~Baer, F.~Quevedo,
G.~Villadoro and L.~Velasco-Sevilla. This work was performed using the
Darwin Supercomputer of the University of Cambridge High Performance
Computing Service (http://www.hpc.cam.ac.uk/), provided by Dell
Inc. using Strategic Research Infrastructure Funding from the Higher
Education Funding Council for England and funding from the Science and
Technology Facilities Council.

\paragraph{Appendix: The formulae for $\Sigma_{u,d}^{u,d}(i), i =
  \tilde{t}_{1,2}, \tilde{b}_{1,2}$} \label{append}  
At one-loop the $\Sigma_{u,d}^{u,d}$ get both particle and sparticle
contributions~\cite{Arnowitt:1992qp, Gladyshev:1996fx} but here we considered only the
$\Sigma_u^u(\tilde t_{1,2}, \tilde b_{1,2})$ ones. For the stops,
\be
\Sigma_u^u(\tilde t_{1,2}) = \frac{3}{16\pi^2} \, F(m_{\tilde
  t_{1,2}}^2) \times \left[ f_t^2 - g_Z^2 \mp \frac{f_t^2
    A_t^2-8g_Z^2(\frac{1}{4}-\frac{2}{3}x_W)\Delta_t}{m_{\tilde t_2}^2
    - m_{\tilde t_1}^2} \right] \textrm{ and }
\ee
\be
\Sigma_d^d (\tilde t_{1,2}) = \frac{3}{16\pi^2} \, F(m_{\tilde
  t_{1,2}}^2) \left[ g_Z^2 \mp \frac{f_t^2\mu^2 + 8 g_Z^2 (\frac{1}{4}
    - \frac{2}{3} x_W) \Delta_t }{m_{\tilde t_2}^2 - m_{\tilde
      t_1}^2}\right]
\ee
where $\Delta_t=(m_{\tilde t_L}^2-m_{\tilde t_R}^2)/2 + M_Z^2 \, \cos
2\beta (\frac{1}{4} - \frac{2}{3} x_W)$, $g_Z^2=(g^2+g^{\prime 2})/8$,
$x_W \equiv \sin^2\theta_W$ and $F(m^2)=m^2\left(\log
(m^2/Q^2)-1\right)$, with $Q^2=m_{\tilde t_1}m_{\tilde
  t_2}$. $m_{\tilde t_{1,2}}$ are computed at tree-level. 
For the bottom-squarks,
\be
\Sigma_u^u (\tilde b_{1,2}) = \frac{3}{16\pi^2} \, F(m_{\tilde
  b_{1,2}}^2) \left[ g_Z^2 \mp \frac{f_b^2 \mu^2 - 8g_Z^2 (\frac{1}{4}
    - \frac{1}{3} x_W) \Delta_b}{m_{\tilde b_2}^2 - m_{\tilde
      b_1}^2}\right]  \textrm{ and } 
\ee
\be
\Sigma_d^d (\tilde b_{1,2}) = \frac{3}{16\pi^2} \, F(m_{\tilde
  b_{1,2}}^2) \left[ f_b^2 - g_Z^2 \mp \frac{f_b^2 A_b^2-8 g_Z^2
    (\frac{1}{4} - \frac{1}{3} x_W) \Delta_b}{m_{\tilde b_2}^2 -
    m_{\tilde b_1}^2}\right] 
\ee
where $\Delta_b=(m_{\tilde b_L}^2-m_{\tilde b_R}^2)/2 + M_Z^2 \, \cos
2\beta (\frac{1}{4} - \frac{1}{3} x_W)$. $m_{\tilde b_{1,2}}$ are
computed at tree-level.

\tiny{
\bibliographystyle{utphys}
\bibliography{Bsmupmum}

\providecommand{\href}[2]{#2}\begingroup\raggedright\begin{thebibliography}{10}

\bibitem{Aad:2012tfa}
{\bfseries ATLAS} Collaboration, G.~Aad {\em et~al.}, ``{Observation of a new
  particle in the search for the Standard Model Higgs boson with the ATLAS
  detector at the LHC},''
  \href{http://dx.doi.org/10.1016/j.physletb.2012.08.020}{{\em Phys.Lett.}
  {\bfseries B716} (2012) 1--29},
\href{http://arxiv.org/abs/1207.7214}{{\ttfamily arXiv:1207.7214 [hep-ex]}}.

\bibitem{Chatrchyan:2012ufa}
{\bfseries CMS} Collaboration, S.~Chatrchyan {\em et~al.}, ``{Observation of a
  new boson at a mass of 125 GeV with the CMS experiment at the LHC},''
  \href{http://dx.doi.org/10.1016/j.physletb.2012.08.021}{{\em Phys.Lett.}
  {\bfseries B716} (2012) 30--61},
\href{http://arxiv.org/abs/1207.7235}{{\ttfamily arXiv:1207.7235 [hep-ex]}}.

\bibitem{AbdusSalam:2008uv}
S.~S. AbdusSalam, ``{The Full 24-Parameter MSSM Exploration},''
  \href{http://dx.doi.org/10.1063/1.3051939}{{\em AIP Conf.Proc.} {\bfseries
  1078} (2009) 297--299},
\href{http://arxiv.org/abs/0809.0284}{{\ttfamily arXiv:0809.0284 [hep-ph]}}.

\bibitem{AbdusSalam:2009qd}
S.~S. AbdusSalam, B.~C. Allanach, F.~Quevedo, F.~Feroz, and M.~Hobson,
  ``{Fitting the Phenomenological MSSM},''
  \href{http://dx.doi.org/10.1103/PhysRevD.81.095012}{{\em Phys.Rev.}
  {\bfseries D81} (2010) 095012},
\href{http://arxiv.org/abs/0904.2548}{{\ttfamily arXiv:0904.2548 [hep-ph]}}.

\bibitem{AbdusSalam:2010qp}
S.~AbdusSalam and F.~Quevedo, ``{Cold Dark Matter Hypotheses in the MSSM},''
  \href{http://dx.doi.org/10.1016/j.physletb.2011.02.065}{{\em Phys.Lett.}
  {\bfseries B700} (2011) 343--350},
\href{http://arxiv.org/abs/1009.4308}{{\ttfamily arXiv:1009.4308 [hep-ph]}}.

\bibitem{AbdusSalam:2011hd}
S.~AbdusSalam, ``{Can the LHC rule out the MSSM?},''
  \href{http://dx.doi.org/10.1016/j.physletb.2011.10.023}{{\em Phys.Lett.}
  {\bfseries B705} (2011) 331--336},
\href{http://arxiv.org/abs/1106.2317}{{\ttfamily arXiv:1106.2317 [hep-ph]}}.

\bibitem{AbdusSalam:2011fc}
S.~AbdusSalam, B.~Allanach, H.~Dreiner, J.~Ellis, U.~Ellwanger, {\em et~al.},
  ``{Benchmark Models, Planes, Lines and Points for Future SUSY Searches at the
  LHC},'' \href{http://dx.doi.org/10.1140/epjc/s10052-011-1835-7}{{\em
  Eur.Phys.J.} {\bfseries C71} (2011) 1835},
\href{http://arxiv.org/abs/1109.3859}{{\ttfamily arXiv:1109.3859 [hep-ph]}}.

\bibitem{AbdusSalam:2012sy}
S.~S. AbdusSalam and D.~Choudhury, ``{Higgs boson discovery versus sparticles
  prediction: Impact on the pMSSM's posterior samples from a Bayesian global
  fit},''
\href{http://arxiv.org/abs/1210.3331}{{\ttfamily arXiv:1210.3331 [hep-ph]}}.

\bibitem{AbdusSalam:2012ir}
S.~S. AbdusSalam, ``{LHC-7 supersymmetry search interpretation within the
  pMSSM},'' \href{http://dx.doi.org/10.1103/PhysRevD.87.115012}{{\em Phys.Rev.}
  {\bfseries D87} (2013) 115012},
\href{http://arxiv.org/abs/1211.0999}{{\ttfamily arXiv:1211.0999 [hep-ph]}}.

\bibitem{deCarlos:1993yy}
B.~de~Carlos and J.~Casas, ``{One loop analysis of the electroweak breaking in
  supersymmetric models and the fine tuning problem},''
  \href{http://dx.doi.org/10.1016/0370-2693(93)90940-J}{{\em Phys.Lett.}
  {\bfseries B309} (1993) 320--328},
\href{http://arxiv.org/abs/hep-ph/9303291}{{\ttfamily arXiv:hep-ph/9303291
  [hep-ph]}}.

\bibitem{Chan:1997bi}
K.~L. Chan, U.~Chattopadhyay, and P.~Nath, ``{Naturalness, weak scale
  supersymmetry and the prospect for the observation of supersymmetry at the
  Tevatron and at the CERN LHC},''
  \href{http://dx.doi.org/10.1103/PhysRevD.58.096004}{{\em Phys.Rev.}
  {\bfseries D58} (1998) 096004},
\href{http://arxiv.org/abs/hep-ph/9710473}{{\ttfamily arXiv:hep-ph/9710473
  [hep-ph]}}.

\bibitem{Dimopoulos:1995mi}
S.~Dimopoulos and G.~Giudice, ``{Naturalness constraints in supersymmetric
  theories with nonuniversal soft terms},''
  \href{http://dx.doi.org/10.1016/0370-2693(95)00961-J}{{\em Phys.Lett.}
  {\bfseries B357} (1995) 573--578},
\href{http://arxiv.org/abs/hep-ph/9507282}{{\ttfamily arXiv:hep-ph/9507282
  [hep-ph]}}.

\bibitem{Berezinsky:1995cj}
V.~Berezinsky, A.~Bottino, J.~R. Ellis, N.~Fornengo, G.~Mignola, {\em et~al.},
  ``{Neutralino dark matter in supersymmetric models with nonuniversal scalar
  mass terms},'' \href{http://dx.doi.org/10.1016/0927-6505(95)00048-8}{{\em
  Astropart.Phys.} {\bfseries 5} (1996) 1--26},
\href{http://arxiv.org/abs/hep-ph/9508249}{{\ttfamily arXiv:hep-ph/9508249
  [hep-ph]}}.

\bibitem{Cassel:2011tg}
S.~Cassel, D.~Ghilencea, S.~Kraml, A.~Lessa, and G.~Ross, ``{Fine-tuning
  implications for complementary dark matter and LHC SUSY searches},''
  \href{http://dx.doi.org/10.1007/JHEP05(2011)120}{{\em JHEP} {\bfseries 1105}
  (2011) 120},
\href{http://arxiv.org/abs/1101.4664}{{\ttfamily arXiv:1101.4664 [hep-ph]}}.

\bibitem{Ghilencea:2012qk}
D.~Ghilencea and G.~Ross, ``{The fine-tuning cost of the likelihood in SUSY
  models},'' \href{http://dx.doi.org/10.1016/j.nuclphysb.2012.11.007}{{\em
  Nucl.Phys.} {\bfseries B868} (2013) 65--74},
\href{http://arxiv.org/abs/1208.0837}{{\ttfamily arXiv:1208.0837 [hep-ph]}}.

\bibitem{Kobayashi:2006fh}
T.~Kobayashi, H.~Terao, and A.~Tsuchiya, ``{Fine-tuning in gauge mediated
  supersymmetry breaking models and induced top Yukawa coupling},''
  \href{http://dx.doi.org/10.1103/PhysRevD.74.015002}{{\em Phys.Rev.}
  {\bfseries D74} (2006) 015002},
\href{http://arxiv.org/abs/hep-ph/0604091}{{\ttfamily arXiv:hep-ph/0604091
  [hep-ph]}}.

\bibitem{Arvanitaki:2012ps}
A.~Arvanitaki, N.~Craig, S.~Dimopoulos, and G.~Villadoro, ``{Mini-Split},''
  \href{http://dx.doi.org/10.1007/JHEP02(2013)126}{{\em JHEP} {\bfseries 1302}
  (2013) 126},
\href{http://arxiv.org/abs/1210.0555}{{\ttfamily arXiv:1210.0555 [hep-ph]}}.

\bibitem{ArkaniHamed:2012gw}
N.~Arkani-Hamed, A.~Gupta, D.~E. Kaplan, N.~Weiner, and T.~Zorawski, ``{Simply
  Unnatural Supersymmetry},''
\href{http://arxiv.org/abs/1212.6971}{{\ttfamily arXiv:1212.6971 [hep-ph]}}.

\bibitem{Feng:2013pwa}
J.~L. Feng, ``{Naturalness and the Status of Supersymmetry},''
  \href{http://dx.doi.org/10.1146/annurev-nucl-102010-130447}{{\em
  Ann.Rev.Nucl.Part.Sci.} {\bfseries 63} (2013) 351--382},
\href{http://arxiv.org/abs/1302.6587}{{\ttfamily arXiv:1302.6587 [hep-ph]}}.

\bibitem{Baer:2012up}
H.~Baer, V.~Barger, P.~Huang, A.~Mustafayev, and X.~Tata, ``{Radiative natural
  SUSY with a 125 GeV Higgs boson},''
  \href{http://dx.doi.org/10.1103/PhysRevLett.109.161802}{{\em Phys.Rev.Lett.}
  {\bfseries 109} (2012) 161802},
\href{http://arxiv.org/abs/1207.3343}{{\ttfamily arXiv:1207.3343 [hep-ph]}}.

\bibitem{Baer:2012cf}
H.~Baer, V.~Barger, P.~Huang, D.~Mickelson, A.~Mustafayev, {\em et~al.},
  ``{Radiative natural supersymmetry: Reconciling electroweak fine-tuning and
  the Higgs boson mass},''
  \href{http://dx.doi.org/10.1103/PhysRevD.87.115028}{{\em Phys.Rev.}
  {\bfseries D87} (2013) 115028},
\href{http://arxiv.org/abs/1212.2655}{{\ttfamily arXiv:1212.2655 [hep-ph]}}.

\bibitem{Baer:2012mv}
H.~Baer, V.~Barger, P.~Huang, D.~Mickelson, A.~Mustafayev, {\em et~al.},
  ``{Post-LHC7 fine-tuning in the mSUGRA/CMSSM model with a 125 GeV Higgs
  boson},'' \href{http://dx.doi.org/10.1103/PhysRevD.87.035017}{{\em Phys.Rev.}
  {\bfseries D87} no.~3, (2013) 035017},
\href{http://arxiv.org/abs/1210.3019}{{\ttfamily arXiv:1210.3019 [hep-ph]}}.

\bibitem{Nucl.Phys.B306.63}
R.~Barbieri and G.~Giudice, ``{Upper Bounds on Supersymmetric Particle
  Masses},'' {\em Nucl.Phys.} {\bfseries B306} (1988) 63.

\bibitem{ellis}
J.~R. Ellis, K.~Enqvist, D.~V. Nanopoulos, and F.~Zwirner, ``{Observables in
  Low-Energy Superstring Models},'' {\em Mod. Phys. Lett.} {\bfseries A1}
  (1986) 57.

\bibitem{Cabrera:2008tj}
M.~Cabrera, J.~Casas, and R.~Ruiz~de Austri, ``{Bayesian approach and
  Naturalness in MSSM analyses for the LHC},''
  \href{http://dx.doi.org/10.1088/1126-6708/2009/03/075}{{\em JHEP} {\bfseries
  0903} (2009) 075},
\href{http://arxiv.org/abs/0812.0536}{{\ttfamily arXiv:0812.0536 [hep-ph]}}.

\bibitem{Ghilencea:2013nxa}
D.~Ghilencea, ``{SUSY naturalness without prejudice},''
\href{http://arxiv.org/abs/1311.6144}{{\ttfamily arXiv:1311.6144 [hep-ph]}}.

\bibitem{Allanach:2001kg}
B.~Allanach, ``{SOFTSUSY: a program for calculating supersymmetric spectra},''
  \href{http://dx.doi.org/10.1016/S0010-4655(01)00460-X}{{\em
  Comput.Phys.Commun.} {\bfseries 143} (2002) 305--331},
\href{http://arxiv.org/abs/hep-ph/0104145}{{\ttfamily arXiv:hep-ph/0104145
  [hep-ph]}}.

\bibitem{Belanger:2008sj}
G.~Belanger, F.~Boudjema, A.~Pukhov, and A.~Semenov, ``{Dark matter direct
  detection rate in a generic model with micrOMEGAs 2.2},''
  \href{http://dx.doi.org/10.1016/j.cpc.2008.11.019}{{\em Comput.Phys.Commun.}
  {\bfseries 180} (2009) 747--767},
\href{http://arxiv.org/abs/0803.2360}{{\ttfamily arXiv:0803.2360 [hep-ph]}}.

\bibitem{Mahmoudi:2007vz}
F.~Mahmoudi, ``{SuperIso: A Program for calculating the isospin asymmetry of B
  to K* gamma in the MSSM},''
  \href{http://dx.doi.org/10.1016/j.cpc.2007.12.006}{{\em Comput.Phys.Commun.}
  {\bfseries 178} (2008) 745--754},
\href{http://arxiv.org/abs/0710.2067}{{\ttfamily arXiv:0710.2067 [hep-ph]}}.

\bibitem{Heinemeyer:2006px}
S.~Heinemeyer, W.~Hollik, D.~Stockinger, A.~Weber, and G.~Weiglein, ``{Precise
  prediction for M(W) in the MSSM},''
  \href{http://dx.doi.org/10.1088/1126-6708/2006/08/052}{{\em JHEP} {\bfseries
  0608} (2006) 052},
\href{http://arxiv.org/abs/hep-ph/0604147}{{\ttfamily arXiv:hep-ph/0604147
  [hep-ph]}}.

\bibitem{Heinemeyer:2007bw}
S.~Heinemeyer, W.~Hollik, A.~Weber, and G.~Weiglein, ``{$Z$ Pole Observables in
  the MSSM},'' \href{http://dx.doi.org/10.1088/1126-6708/2008/04/039}{{\em
  JHEP} {\bfseries 0804} (2008) 039},
\href{http://arxiv.org/abs/0710.2972}{{\ttfamily arXiv:0710.2972 [hep-ph]}}.

\bibitem{verzo}
M.~Verzocchi in {\em {talk at ICHEP 2008}}.
\newblock 2008, Philadelphia, USA.

\bibitem{:2005ema}
{\bfseries ALEPH} Collaboration, ``{Precision electroweak measurements on the
  $Z$ resonance},'' \href{http://dx.doi.org/10.1016/j.physrep.2005.12.006}{{\em
  Phys. Rept.} {\bfseries 427} (2006) 257},
\href{http://arxiv.org/abs/hep-ex/0509008}{{\ttfamily arXiv:hep-ex/0509008}}.

\bibitem{Bennett:2006fi}
{\bfseries Muon G-2} Collaboration, G.~Bennett {\em et~al.}, ``{Final Report of
  the Muon E821 Anomalous Magnetic Moment Measurement at BNL},''
  \href{http://dx.doi.org/10.1103/PhysRevD.73.072003}{{\em Phys.Rev.}
  {\bfseries D73} (2006) 072003},
\href{http://arxiv.org/abs/hep-ex/0602035}{{\ttfamily arXiv:hep-ex/0602035
  [hep-ex]}}.

\bibitem{Davier:2007ua}
M.~Davier, ``{The Hadronic contribution to (g-2)(mu)},''
  \href{http://dx.doi.org/10.1016/j.nuclphysbps.2007.03.023}{{\em
  Nucl.Phys.Proc.Suppl.} {\bfseries 169} (2007) 288--296},
\href{http://arxiv.org/abs/hep-ph/0701163}{{\ttfamily arXiv:hep-ph/0701163
  [hep-ph]}}.

\bibitem{Barberio:2007cr}
{\bfseries Heavy Flavor Averaging Group (HFAG)} Collaboration, E.~Barberio {\em
  et~al.}, ``{Averages of $b-$hadron properties at the end of 2006},''
\href{http://arxiv.org/abs/0704.3575}{{\ttfamily arXiv:0704.3575 [hep-ex]}}.

\bibitem{Aaij:2012nna}
{\bfseries LHCb} Collaboration, R.~Aaij {\em et~al.}, ``{First Evidence for the
  Decay $B^0_s \to \mu^+\mu^-$},''
  \href{http://dx.doi.org/10.1103/PhysRevLett.110.021801}{{\em Phys.Rev.Lett.}
  {\bfseries 110} (2013) 021801},
\href{http://arxiv.org/abs/1211.2674}{{\ttfamily arXiv:1211.2674 [hep-ex]}}.

\bibitem{Abulencia:2006ze}
{\bfseries CDF} Collaboration, A.~Abulencia {\em et~al.}, ``{Observation of
  B0(s) - anti-B0(s) Oscillations},''
  \href{http://dx.doi.org/10.1103/PhysRevLett.97.242003}{{\em Phys.Rev.Lett.}
  {\bfseries 97} (2006) 242003},
\href{http://arxiv.org/abs/hep-ex/0609040}{{\ttfamily arXiv:hep-ex/0609040
  [hep-ex]}}.

\bibitem{Aubert:2004kz}
{\bfseries BaBar} Collaboration, B.~Aubert {\em et~al.}, ``{Search for the rare
  leptonic decay $B^- \to \tau^- \bar{\nu}_\tau$},''
  \href{http://dx.doi.org/10.1103/PhysRevLett.95.041804}{{\em Phys.Rev.Lett.}
  {\bfseries 95} (2005) 041804},
\href{http://arxiv.org/abs/hep-ex/0407038}{{\ttfamily arXiv:hep-ex/0407038
  [hep-ex]}}.

\bibitem{paoti}
P.~Chang in {\em {talk at ICHEP 2008}}.
\newblock 2008, Philadelphia, USA.

\bibitem{hep-lat/0507015}
{\bfseries HPQCD} Collaboration, A.~Gray {\em et~al.}, ``{The B meson decay
  constant from unquenched lattice QCD},''
  \href{http://dx.doi.org/10.1103/PhysRevLett.95.212001}{{\em Phys.Rev.Lett.}
  {\bfseries 95} (2005) 212001},
\href{http://arxiv.org/abs/hep-lat/0507015}{{\ttfamily arXiv:hep-lat/0507015
  [hep-lat]}}.

\bibitem{J.Phys.G33.1}
{\bfseries Particle Data Group} Collaboration, C.~Amsler {\em et~al.},
  ``{Review of particle physics},''
\href{http://dx.doi.org/10.1016/j.physletb.2008.07.018}{{\em Phys. Lett.}
  {\bfseries B667} (2008) 1}.

\bibitem{0803.0547}
{\bfseries WMAP} Collaboration, E.~Komatsu {\em et~al.}, ``{Five-Year Wilkinson
  Microwave Anisotropy Probe (WMAP) Observations: Cosmological
  Interpretation},'' \href{http://dx.doi.org/10.1088/0067-0049/180/2/330}{{\em
  Astrophys.J.Suppl.} {\bfseries 180} (2009) 330--376},
\href{http://arxiv.org/abs/0803.0547}{{\ttfamily arXiv:0803.0547 [astro-ph]}}.

\bibitem{ATLAS:2013mma}
{\bfseries ATLAS} Collaboration, ``{Combined measurements of the mass and
  signal strength of the Higgs-like boson with the ATLAS detector using up to
  25 fb$^{-1}$ of proton-proton collision data},''
\href{http://arxiv.org/abs/ATLAS-CONF-2013-014,
  ATLAS-COM-CONF-2013-025}{{\ttfamily ATLAS-CONF-2013-014,
  ATLAS-COM-CONF-2013-025}}.

\bibitem{CMS:yva}
{\bfseries CMS} Collaboration, ``{Combination of standard model Higgs boson
  searches and measurements of the properties of the new boson with a mass near
  125 GeV},''
\href{http://arxiv.org/abs/CMS-PAS-HIG-13-005}{{\ttfamily CMS-PAS-HIG-13-005}}.

\bibitem{Feroz:2007kg}
F.~Feroz and M.~Hobson, ``{Multimodal nested sampling: an efficient and robust
  alternative to MCMC methods for astronomical data analysis},''
  \href{http://dx.doi.org/10.1111/j.1365-2966.2007.12353.x}{{\em
  Mon.Not.Roy.Astron.Soc.} {\bfseries 384} (2008) 449},
\href{http://arxiv.org/abs/0704.3704}{{\ttfamily arXiv:0704.3704 [astro-ph]}}.

\bibitem{Feroz:2008xx}
F.~Feroz, M.~Hobson, and M.~Bridges, ``{MultiNest: an efficient and robust
  Bayesian inference tool for cosmology and particle physics},''
  \href{http://dx.doi.org/10.1111/j.1365-2966.2009.14548.x}{{\em
  Mon.Not.Roy.Astron.Soc.} {\bfseries 398} (2009) 1601--1614},
\href{http://arxiv.org/abs/0809.3437}{{\ttfamily arXiv:0809.3437 [astro-ph]}}.

\bibitem{Skilling}
J.~{Skilling}, \href{http://dx.doi.org/10.1063/1.1835238}{``{Nested
  Sampling},''} in {\em American Institute of Physics Conference Series},
  R.~{Fischer}, R.~{Preuss}, and U.~V. {Toussaint}, eds., pp.~395--405.
\newblock Nov., 2004.
\newblock \url{http://www.inference.phy.cam.ac.uk/bayesys/}.

\bibitem{ATLAS:2012rov}
{\bfseries ATLAS} Collaboration, ``{Search for a supersymmetric top-quark
  partner in final states with two leptons in sqrt(s) = 8 TeV pp collisions
  using 13 fb$^{-1}$ of ATLAS data},''
\href{http://arxiv.org/abs/ATLAS-CONF-2012-167,
  ATLAS-COM-CONF-2012-202}{{\ttfamily ATLAS-CONF-2012-167,
  ATLAS-COM-CONF-2012-202}}.

\bibitem{ATLAS:2013pla}
{\bfseries ATLAS} Collaboration, ``{Search for direct top squark pair
  production in final states with one isolated lepton, jets, and missing
  transverse momentum in $sqrt{s}=8,$TeV $pp$ collisions using 21 fb$^{-1}$ of
  ATLAS data},''
\href{http://arxiv.org/abs/ATLAS-CONF-2013-037,
  ATLAS-COM-CONF-2013-038}{{\ttfamily ATLAS-CONF-2013-037,
  ATLAS-COM-CONF-2013-038}}.

\bibitem{Aad:2012ywa}
{\bfseries ATLAS} Collaboration, G.~Aad {\em et~al.}, ``{Search for a
  supersymmetric partner to the top quark in final states with jets and missing
  transverse momentum at $\sqrt{s}=7$ TeV with the ATLAS detector},''
  \href{http://dx.doi.org/10.1103/PhysRevLett.109.211802}{{\em Phys.Rev.Lett.}
  {\bfseries 109} (2012) 211802},
\href{http://arxiv.org/abs/1208.1447}{{\ttfamily arXiv:1208.1447 [hep-ex]}}.

\bibitem{Aad:2012xqa}
{\bfseries ATLAS} Collaboration, G.~Aad {\em et~al.}, ``{Search for direct top
  squark pair production in final states with one isolated lepton, jets, and
  missing transverse momentum in $\sqrt{s}=7$ TeV $pp$ collisions using 4.7
  $fb^{-1}$ of ATLAS data},''
  \href{http://dx.doi.org/10.1103/PhysRevLett.109.211803}{{\em Phys.Rev.Lett.}
  {\bfseries 109} (2012) 211803},
\href{http://arxiv.org/abs/1208.2590}{{\ttfamily arXiv:1208.2590 [hep-ex]}}.

\bibitem{Aad:2012tx}
{\bfseries ATLAS} Collaboration, G.~Aad {\em et~al.}, ``{Search for light
  scalar top quark pair production in final states with two leptons with the
  ATLAS detector in $\sqrt{s}=7$ TeV proton-proton collisions},''
  \href{http://dx.doi.org/10.1140/epjc/s10052-012-2237-1}{{\em Eur.Phys.J.}
  {\bfseries C72} (2012) 2237},
\href{http://arxiv.org/abs/1208.4305}{{\ttfamily arXiv:1208.4305 [hep-ex]}}.

\bibitem{Aad:2012uu}
{\bfseries ATLAS} Collaboration, G.~Aad {\em et~al.}, ``{Search for a heavy
  top-quark partner in final states with two leptons with the ATLAS detector at
  the LHC},'' \href{http://dx.doi.org/10.1007/JHEP11(2012)094}{{\em JHEP}
  {\bfseries 1211} (2012) 094},
\href{http://arxiv.org/abs/1209.4186}{{\ttfamily arXiv:1209.4186 [hep-ex]}}.

\bibitem{Aad:2012yr}
{\bfseries ATLAS} Collaboration, G.~Aad {\em et~al.}, ``{Search for light top
  squark pair production in final states with leptons and $b^-$ jets with the
  ATLAS detector in $\sqrt{s}=7$ TeV proton-proton collisions},''
  \href{http://dx.doi.org/10.1016/j.physletb.2013.01.049}{{\em Phys.Lett.}
  {\bfseries B720} (2013) 13--31},
\href{http://arxiv.org/abs/1209.2102}{{\ttfamily arXiv:1209.2102 [hep-ex]}}.

\bibitem{ATLAS:2013cma}
{\bfseries ATLAS} Collaboration, ``{Search for direct production of the top
  squark in the all-hadronic ttbar + etmiss final state in 21 fb-1 of
  p-pcollisions at sqrt(s)=8 TeV with the ATLAS detector},''
\href{http://arxiv.org/abs/ATLAS-CONF-2013-024,
  ATLAS-COM-CONF-2013-011}{{\ttfamily ATLAS-CONF-2013-024,
  ATLAS-COM-CONF-2013-011}}.

\bibitem{TheATLAScollaboration:2013gha}
{\bfseries ATLAS} Collaboration, ``{Search for direct top squark pair
  production in final states with two leptons in $\sqrt{s}$ = 8 TeV pp
  collisions using $20$fb$^{-1}$ of ATLAS data.},''
\href{http://arxiv.org/abs/ATLAS-CONF-2013-048,
  ATLAS-COM-CONF-2013-056}{{\ttfamily ATLAS-CONF-2013-048,
  ATLAS-COM-CONF-2013-056}}.

\bibitem{TheATLAScollaboration:2013xha}
{\bfseries ATLAS} Collaboration, ``{Searches for direct scalar top pair
  production in final states with two leptons using the stransverse mass
  variable and a multivariate analysis technique in $\sqrt{s} = 8$ TeV pp
  collisions using 20.3 fb$^{-1}$ of ATLAS data},''
\href{http://arxiv.org/abs/ATLAS-CONF-2013-065,
  ATLAS-COM-CONF-2013-065}{{\ttfamily ATLAS-CONF-2013-065,
  ATLAS-COM-CONF-2013-065}}.

\bibitem{TheATLAScollaboration:2013aia}
{\bfseries ATLAS} Collaboration, ``{Search for pair-produced top squarks
  decaying into a charm quark and the lightest neutralinos with 20.3
  fb${}^{-1}$ of $pp$ collisions at $\sqrt{s}=8~$TeV with the ATLAS detector at
  the LHC},''
\href{http://arxiv.org/abs/ATLAS-CONF-2013-068,
  ATLAS-COM-CONF-2013-076}{{\ttfamily ATLAS-CONF-2013-068,
  ATLAS-COM-CONF-2013-076}}.

\bibitem{Chatrchyan:2013lya}
{\bfseries CMS} Collaboration, S.~Chatrchyan {\em et~al.}, ``{Search for
  supersymmetry in hadronic final states with missing transverse energy using
  the variables AlphaT and b-quark multiplicity in pp collisions at 8 TeV},''
  \href{http://dx.doi.org/10.1140/epjc/s10052-013-2568-6}{{\em Eur.Phys.J.}
  {\bfseries C73} (2013) 2568},
\href{http://arxiv.org/abs/1303.2985}{{\ttfamily arXiv:1303.2985 [hep-ex]}}.

\bibitem{CMS:2013cfa}
{\bfseries CMS} Collaboration, CMS, ``{Search for supersymmetry using razor
  variables in events with b-jets in pp collisions at 8 TeV},''
\href{http://arxiv.org/abs/CMS-PAS-SUS-13-004}{{\ttfamily CMS-PAS-SUS-13-004}}.

\bibitem{CMS:2013hda}
{\bfseries CMS} Collaboration, CMS, ``{Search for top-squark pair production in
  the single lepton final state in pp collisions at 8 TeV},''
\href{http://arxiv.org/abs/CMS-PAS-SUS-13-011}{{\ttfamily CMS-PAS-SUS-13-011}}.

\bibitem{Aad:2013yna}
{\bfseries ATLAS} Collaboration, G.~Aad {\em et~al.}, ``{Search for charginos
  nearly mass-degenerate with the lightest neutralino based on a
  disappearing-track signature in pp collisions at $\sqrt{s}$ = 8 TeV with the
  ATLAS detector},'' \href{http://dx.doi.org/10.1103/PhysRevD.88.112006}{{\em
  Phys.Rev.} {\bfseries D88} (2013) 112006},
\href{http://arxiv.org/abs/1310.3675}{{\ttfamily arXiv:1310.3675 [hep-ex]}}.

\bibitem{Heister:2002mn}
{\bfseries ALEPH} Collaboration, A.~Heister {\em et~al.}, ``{Search for
  charginos nearly mass degenerate with the lightest neutralino in e+ e-
  collisions at center-of-mass energies up to 209-GeV},''
  \href{http://dx.doi.org/10.1016/S0370-2693(02)01584-8}{{\em Phys.Lett.}
  {\bfseries B533} (2002) 223--236},
\href{http://arxiv.org/abs/hep-ex/0203020}{{\ttfamily arXiv:hep-ex/0203020
  [hep-ex]}}.

\bibitem{Acciarri:2000wy}
{\bfseries L3} Collaboration, M.~Acciarri {\em et~al.}, ``{Search for charginos
  with a small mass difference with the lightest supersymmetric particle at
  $\sqrt{S}$ = 189-GeV},''
  \href{http://dx.doi.org/10.1016/S0370-2693(00)00488-3}{{\em Phys.Lett.}
  {\bfseries B482} (2000) 31--42},
\href{http://arxiv.org/abs/hep-ex/0002043}{{\ttfamily arXiv:hep-ex/0002043
  [hep-ex]}}.

\bibitem{Abbiendi:2002vz}
{\bfseries OPAL} Collaboration, G.~Abbiendi {\em et~al.}, ``{Search for nearly
  mass degenerate charginos and neutralinos at LEP},''
  \href{http://dx.doi.org/10.1140/epjc/s2003-01237-x}{{\em Eur.Phys.J.}
  {\bfseries C29} (2003) 479--489},
\href{http://arxiv.org/abs/hep-ex/0210043}{{\ttfamily arXiv:hep-ex/0210043
  [hep-ex]}}.

\bibitem{ElKheishen:1992yv}
M.~El~Kheishen, A.~Aboshousha, and A.~Shafik, ``{Analytic formulas for the
  neutralino masses and the neutralino mixing matrix},''
\href{http://dx.doi.org/10.1103/PhysRevD.45.4345}{{\em Phys.Rev.} {\bfseries
  D45} (1992) 4345--4348}.

\bibitem{Bottino:2000jx}
A.~Bottino, F.~Donato, N.~Fornengo, and S.~Scopel, ``{Probing the
  supersymmetric parameter space by WIMP direct detection},''
  \href{http://dx.doi.org/10.1103/PhysRevD.63.125003}{{\em Phys.Rev.}
  {\bfseries D63} (2001) 125003},
\href{http://arxiv.org/abs/hep-ph/0010203}{{\ttfamily arXiv:hep-ph/0010203
  [hep-ph]}}.

\bibitem{Akerib:2005za}
{\bfseries CDMS} Collaboration, D.~Akerib {\em et~al.}, ``{Limits on
  spin-dependent wimp-nucleon interactions from the cryogenic dark matter
  search},'' \href{http://dx.doi.org/10.1103/PhysRevD.73.011102}{{\em
  Phys.Rev.} {\bfseries D73} (2006) 011102},
\href{http://arxiv.org/abs/astro-ph/0509269}{{\ttfamily arXiv:astro-ph/0509269
  [astro-ph]}}.

\bibitem{Aprile:2012nq}
{\bfseries XENON100} Collaboration, E.~Aprile {\em et~al.}, ``{Dark Matter
  Results from 225 Live Days of XENON100 Data},''
  \href{http://dx.doi.org/10.1103/PhysRevLett.109.181301}{{\em Phys.Rev.Lett.}
  {\bfseries 109} (2012) 181301},
\href{http://arxiv.org/abs/1207.5988}{{\ttfamily arXiv:1207.5988
  [astro-ph.CO]}}.

\bibitem{Akerib:2013tjd}
{\bfseries LUX} Collaboration, D.~Akerib {\em et~al.}, ``{First results from
  the LUX dark matter experiment at the Sanford Underground Research
  Facility},''
\href{http://arxiv.org/abs/1310.8214}{{\ttfamily arXiv:1310.8214
  [astro-ph.CO]}}.

\bibitem{Aprile:2012zx}
{\bfseries XENON1T collaboration} Collaboration, E.~Aprile, ``{The XENON1T Dark
  Matter Search Experiment},''
\href{http://arxiv.org/abs/1206.6288}{{\ttfamily arXiv:1206.6288
  [astro-ph.IM]}}.

\bibitem{Arnowitt:1992qp}
R.~L. Arnowitt and P.~Nath, ``{Loop corrections to radiative breaking of
  electroweak symmetry in supersymmetry},''
\href{http://dx.doi.org/10.1103/PhysRevD.46.3981}{{\em Phys.Rev.} {\bfseries
  D46} (1992) 3981--3986}.

\bibitem{Gladyshev:1996fx}
A.~Gladyshev, D.~Kazakov, W.~de~Boer, G.~Burkart, and R.~Ehret, ``{MSSM
  predictions of the neutral Higgs boson masses and LEP-2 production
  cross-sections},''
  \href{http://dx.doi.org/10.1016/S0550-3213(97)00253-8}{{\em Nucl.Phys.}
  {\bfseries B498} (1997) 3--27},
\href{http://arxiv.org/abs/hep-ph/9603346}{{\ttfamily arXiv:hep-ph/9603346
  [hep-ph]}}.

\end{thebibliography}\endgroup
}
\end{document}